\documentstyle[epsfig]{aipproc}
\begin{document}
\title{The Mystery of Flavor}
\author{R. D. Peccei}
\address{Department of Physics and Astronomy, UCLA, Los Angeles, CA 
90095-1547}
\maketitle
\begin{abstract}
After outlining some of the issues surrounding the flavor problem,
I present three speculative ideas on the origin of families.  In turn,
families are conjectured to arise from an underlying preon dynamics;
from random dynamics at very short distances; or as a result of 
compactification in higher dimensional theories.  Examples and
limitations of each of these speculative scenarios are discussed.
The twin roles that family symmetries and GUTs can have on the spectrum
of quarks and leptons is emphasized, along with the dominant role that
the top mass is likely to play in the dynamics of mass generation.
\end{abstract}

\section{The Question of Flavor}

Flavor is an old problem.  I. I. Rabi's famous question about the muon:
``who ordered that?" has now been replaced by an equally difficult
question to answer: ``why do we have three families of quarks and
leptons?"  Although 
qualitatively we understand the issues connected to flavor a lot
better now, quantitatively we are as puzzled as when the muon
was discovered.

When thinking of flavor, it is useful to consider the standard model
Lagrangian in a sequence of steps.  At the roughest level, neglecting both
gauge and Yukawa interactions, the Standard Model Lagrangian
${\cal{L}}_o = {\cal{L}}_{\rm SM}(g_i=0;\Gamma_{ij}=0)$ has a $U(48)$ 
global symmetry corresponding to the freedom of being able to interchange
any of the 16 fermions of the 3 families of quarks and leptons with
one another.  If we turn on the gauge interactions, the
Lagrangian ${\cal{L}}_1 = {\cal{L}}_{\rm SM}(g_i \not= 0;\Gamma_{ij}=0)$
has a much more restricted symmetry $[U(3))]^6$ corresponding to 
interchanging fermions of a given type (e.g. the $(u,d)_{\rm L}$
doublet)
from one family to the other.  When also the Yukawa interactions are turned
on, ${\cal{L}}_2 = {\cal{L}}_{\rm SM}(g_i\not= 0;\Gamma_{ij}\not= 0)$, then
the only remaining symmetry of the Lagrangian is $U(1)_{\rm B}\times
U(1)_{\rm L}$.  In fact, because of the chiral anomaly\cite{ABJ}, at the
quantum level the symmetry of ${\cal{L}}_2$ is just $U(1)_{\rm B-L}$.

The above classification scheme serves to emphasize that there are really
three distinct flavor problems.  There is a {\bf matter problem }, a
{\bf family problem} and a {\bf mass problem}.  The first of these
problems is simply that of understanding the origin of the different
species of quarks and leptons (i.e. why does one have a $\nu^c_{\rm L}$
and a $u^c_{\rm L}$ state?).  The second problem is related to the 
triplication of the quarks and leptons.  What physics forces such a 
triplication?  Finally, the last problem is related to understanding the origin
of the observed peculiar mass pattern of the known fermions.

The usual approach when thinking about flavor is to try to decouple the
above three problems from one another.  Thus, for example, one assumes the
existence of the quarks and leptons in the Standard Model and asks for
the physics behind the replication of families.  Although it is difficult
to argue cogently on this point, it is certainly true in the examples 
which we will discuss that the
matter problem seems to be unrelated to the question of family replication.
Indeed, 
quite often one also assumes the reverse, namely, that the family replication
question is independent of the types of quarks and leptons one has.  
In fact, it is
possible that there is other matter besides the known quarks and
leptons and that this matter is also replicated.  Certainly, even in the
minimal
Standard Model there is {\bf other matter} besides
the quarks and leptons, connected to the symmetry breaking sector.  
This raises a host of questions including that of 
possible family replication of the ordinary Higgs doublet.  One knows,
empirically, that this cannot happen if one is to avoid flavor changing neutral
currents (FCNC)\cite{GW}.  
However, some replication is needed if there is supersymmetry,
but the 
two different 
Higgs doublets needed in supersymmetry
are connected with different quark charges and need not
replicate as families.

The above remarks suggests that there are some perils associated with trying to seek the origin for family replication independently from that of the quarks
and leptons themselves.  Nevertheless, that is the approach usually taken and the one I will follow here.  Similarly, one also usually tries to disconnect
the problem of mass from that of matter and family.  That is, one generally
assumes the existence of the three observed families of quarks and leptons, 
and then tries to postulate (approximate) symmetries of the mass matrices
for quarks and leptons which will give interrelations among the masses
and mixing parameters for some of these states.

This approach usually involves some kind of {\bf family symmetry} and
is sensible provided that:
\begin{description}
\item{i)} There is some misalignment between the mass matrix basis and the
gauge interaction basis for the quarks and leptons.  Only through such a
misalignment will there result a nontrivial mixing matrix: $V_{\rm CKM}
\not= 1$.
\item{ii)} The family symmetries of the mass matrices are broken
(otherwise \break $V_{CKM} \equiv 1$) either explicitly or spontaneously.
Furthermore, if the breaking is spontaneous, it must occur at a
sufficiently high scale to have escaped detection so far.
\end{description}

Although the origin of flavor remains a mystery, I want to
discuss here three speculative ideas for the origin of families.  These ideas
are realized up to now only in incomplete ways, in what amount 
essentially to toy models.
Thus, for instance, the issue of family generation is in general
disconnected from the question of $SU(2)\times U(1)$ breaking and, often,
also from trying to explicitly calculate the Yukawa couplings.  As a result,
in all of these attempts at trying to understand flavor, the question
of mass is approached from a much more phenomenological viewpoint.  One
guesses certain family or GUT symmetries, and their possible patterns of
breaking, and then one checks out these guesses by testing their predictions
experimentally.  In all of these considerations, 
the top mass, because it is the
dominant mass in the spectrum, plays a fundamental role. 

In my lectures \cite{RDP}, I will begin by describing
three speculative ideas for the origin of families. Specifically,
I will consider in turn the generation of families dynamically; through
short distance chaotic dynamics; and as a result of geometry.  After this
speculative tour, I will discuss briefly the issue of mass generation. In particular, I will illustrate the twin roles that family symmetries and GUTs can have for the spectrum of quarks and leptons. I will conclude by commenting on the profound role that the top mass is likely to have on the detailed dynamics of mass generation.

\section{Generating Families Dynamically}

The underlying idea behind this approach to the flavor problem is that familiies
of quarks and leptons result because they are themselves composites of yet
more fundamental ingredients--{\bf preons}.  There is a nice isotope analogy\cite{KLS}
which serves to illustrate this point.  Think of the three isotopes
of Hydrogen as three distinct families.  Just like the 
families of quarks and leptons,
all three isotopes have the same interactions--their chemistry being
determined by the electromagnetic interactions of the proton.  Deuterium and tritium, however, have different masses 
than the proton because they have, respectively, 1
and 2 neutrons.  Of course, the analogy is not perfect since $^1H$ and $^3H$
are fermions and $^2H$ is a boson!  Nevertheless, it is tempting to
suppose that the 3 families of quarks and leptons, just like the Hydrogen
isotopes, result from the presence of different ``neutral" constituents.

I will illustrate how to generate families dynamically by using as an 
example some recent work of Kaplan, Lepeintre and Schmaltz\cite{KLS}.
By using essentially the isotope analogy, these authors
constructed an interesting toy
model of flavor.  Their simplest toy
model is based on an underlying supersymmetric gauge theory based on the 
symplectic group $Sp(6)$.  The fundamental constituents
in this model are 6 preons $Q_\alpha$
transforming according to the fundamental representation of $Sp(6)$ and
one preon $A_{\alpha\beta}$ transforming according to the 2-rank
antisymmetric representation.  Such a theory has three families of bound 
states distinguished by their $A_{\alpha\beta}$ content, plus a pair of
(neutral) exotic states.  To wit, the bound states of the model are the
15 flavor states
\begin{equation} 
F_3^{[i,j]} \sim Q^i_\alpha Q^j_\alpha;~~
F_2^{[i,j]} \sim Q^i_\alpha A_{\alpha\beta}Q_\beta^j;~~
F_1^{[i,j]} \sim Q^i_\alpha A_{\alpha\beta} 
A_{\beta\gamma}Q_\gamma^j
\end{equation}
plus the two neutral exotic states
\begin{equation}
T_2\sim {\rm Tr} A^2~; ~~~ T_3 \sim {\rm Tr} A^3~.
\end{equation}
The six $Q_\alpha$ preons act as the protons in the isotope analogy.  In
principle, one could imagine having the $SU(3)\times SU(2)\times U(1)$
interactions act on the $Q_\alpha$ states, while   
the $A_{\alpha\beta}$ preons act as the neutrons.
Furthermore, there is clearly a family $U(1)_{\rm F}$ in the
spectrum which counts the number of $A_{\alpha\beta}$ fields.  Finally, one
should note that, because of the supersymmetry, each of the states in Eqs. (1)
and (2) contain both fermions and bosons.

Although the number of bound states per family (15) is encouraging, these
states cannot really be the ordinary quarks and leptons (minus the
right-handed neutrinos). It turns out that 
one cannot properly incorporate the
$SU(3)\times SU(2)\times U(1)$ gauge interactions with only 6 $Q_\alpha$
preons.  To do that, in fact, one has to at least {\bf triplicate} the
underlying gauge theory\cite{KLS} from $Sp(6)$ to $Sp(6)_{\rm L} \times
Sp(6)_{\rm R}\times Sp(6)_{\rm H}$.  Each of these $Sp(6)$ groups has again
six $Q_\alpha$ and one $A_{\alpha\beta}$ preon.  To obtain 
the desired quarks and leptons 
the $Q$ preons are assumed to have the
following $SU(3)\times SU(2)\times U(1)$ assignments:
\begin{eqnarray}
& Q_{\rm L}: & (3,1)_0 \oplus (1,2)_{1/6} 
\oplus (1,1)_{-1/3} \nonumber \\
& Q_{\rm R}: & (\bar 3,1)_0 \oplus (1,1)_{-2/3}
\oplus 2(1,1)_{1/3}\nonumber \\
& Q_{\rm H}: & (1,2)_{-1/6} \oplus (1,1)_{1/3} \oplus (1,1)_{2/3}
\oplus 2(1,1)_{-1/3} 
\end{eqnarray}

Because of the preon group triplication, instead of having 15
$F^{[i,j]}$ bound states per family, one now has 45 such states.  
Per family, these
states now include 16 states with the quantum numbers of the observed
quarks and leptons, plus 29 exotic states which, however, sit in vector-like
representations of the Standard Model group.  Specifically, the quark
doublet $(u,d)_{\rm L}$ is a bound state of $Sp(6)_{\rm L}$;
$u^c_{\rm L}$ and $d^c_{\rm L}$ are bound states of $Sp(6)_{\rm R}$;
while the lepton states $(\nu,e)_{\rm L}$, $\nu^c_{\rm L}$ and $e^c_{\rm L}$
are bound states of $Sp(6)_{\rm H}$.  Among the exotic states one finds as
bound states of $Sp(6)_{\rm H}$ two states with the quantum numbers of the
Higgs doublets of a supersymmetric theory: $H_1\sim (h_1^o,h_1^-)$ and
$H_2\sim (h_2^+,h_2^o)$.  So, in this model, there is a natural family
repetition of the Higgs states. Naively, this could cause problems with
FCNC.  It turns out, however, that when one calculates the dynamical
superpotential of the theory\cite{Seiberg} one can show\cite{KLS} that
there is a ground state where only one of the three families of Higgs
states are left light.  So, in fact, there are no FCNC problems.

This nice result is tempered by other troublesome features of the model
which render it unrealistic--but not uninteresting.  For example,
to break the $[U_{\rm F}(1)]^3$ family symmetry of the model, it is
necessary to introduce by hand some heavy fields (with masses 
$\mu > \Lambda$--the dynamical scale of the preon theories) which serve to
couple the preon groups together.  The simplest possibility is afforded by
having 3 such fields: $v^1_{\alpha_{\rm H}\beta_{\rm R}},~
v^2_{\alpha_{\rm R}\beta_{\rm L}},~ v^3_{\alpha_{\rm L}\beta_{\rm H}}$
with indices spanning 2 of the preon groups, interacting through a 
superpotential
\begin{eqnarray}
W &=& av^1v^2v^3 + b^1_{\rm H}v^1v^1 A_{\rm H} +
b^1_{\rm R} v^1v^1 A_{\rm R} + b^2_{\rm R} v^2v^2 A_{\rm R} +
b^2_{\rm L}v^2v^2A_{\rm L}\nonumber \\
& & \mbox{}+b^3_{\rm L}v^3v^3 A_{\rm L} +
b^3_{\rm H}v^3v^3 A_{\rm H}
\end{eqnarray}
The $a$-term above ties the preon theories together, while the various
$b$-terms serve to break the family symmetries.  Although Eq. (4) is
introduced by hand, integrating out the effects of the heavy $v^i$ fields
gives effective Yukawa couplings of different strengths, much in the way 
originally suggested by Froggatt and Nielsen\cite{FN}.  This is
illustrated schematically in Fig. 1 for the Yukawa coupling of $u^c_{\rm L}$
with $(c,s)_{\rm L}$ via the Higgs state $H^{(3)}_2$ of the third
family--which is the only one which is assumed to get a VEV.\footnote{In the
model\cite{KLS} the lightest family has the most $A_{\alpha\beta}$ fields--c.f.
Eq. (1).}  
One finds\cite{KLS}
\begin{equation}
\Gamma^{(3)}_{12} \sim a~b_{\rm R}^1b_{\rm R}^2b_{\rm L}^3
(\Lambda/\mu)^6 \sim \epsilon^6
\end{equation}

Although the various elements in the up-and-down quark mass matrices are
hierarchial, unfortunately 
there is no resulting quark mixing since $M_u\sim M_d$.  This follows 
because
the model has an unbroken global $SU(2)$ symmetry at the preon level
corresponding to the interchange of the $(1,1)_{-2/3}$ and
$(1,1)_{1/3}$ assignments in Eq. (3).  Furthermore, for the lepton sectors
there is a dynamically generated set of Yukawa couplings\cite{Seiberg}
which are typically unsuppressed.  As a result, naively, one expects
$m_\tau\gg m_t$.  Both of these results make the $[Sp(6)]^3$ model as
presented above unrealistic.  By further complicating the model, Kaplan,
Lepeintre and Schmaltz\cite{KLS} are able to obtain both a non-trivial CKM
matrix and re-establish the top as the heaviest bound state.  However, these
``improved" models are not particularly attractive and represent, more
than anything else, a ``proof of principle".  In addition, even 
after these
problems are resolved, the models still lack mechanisms for breaking
$SU(2)\times U(1)$ and supersymmetry, features which must be understood to
make contact with reality.
\begin{figure}
\begin{center}
\epsfig{file=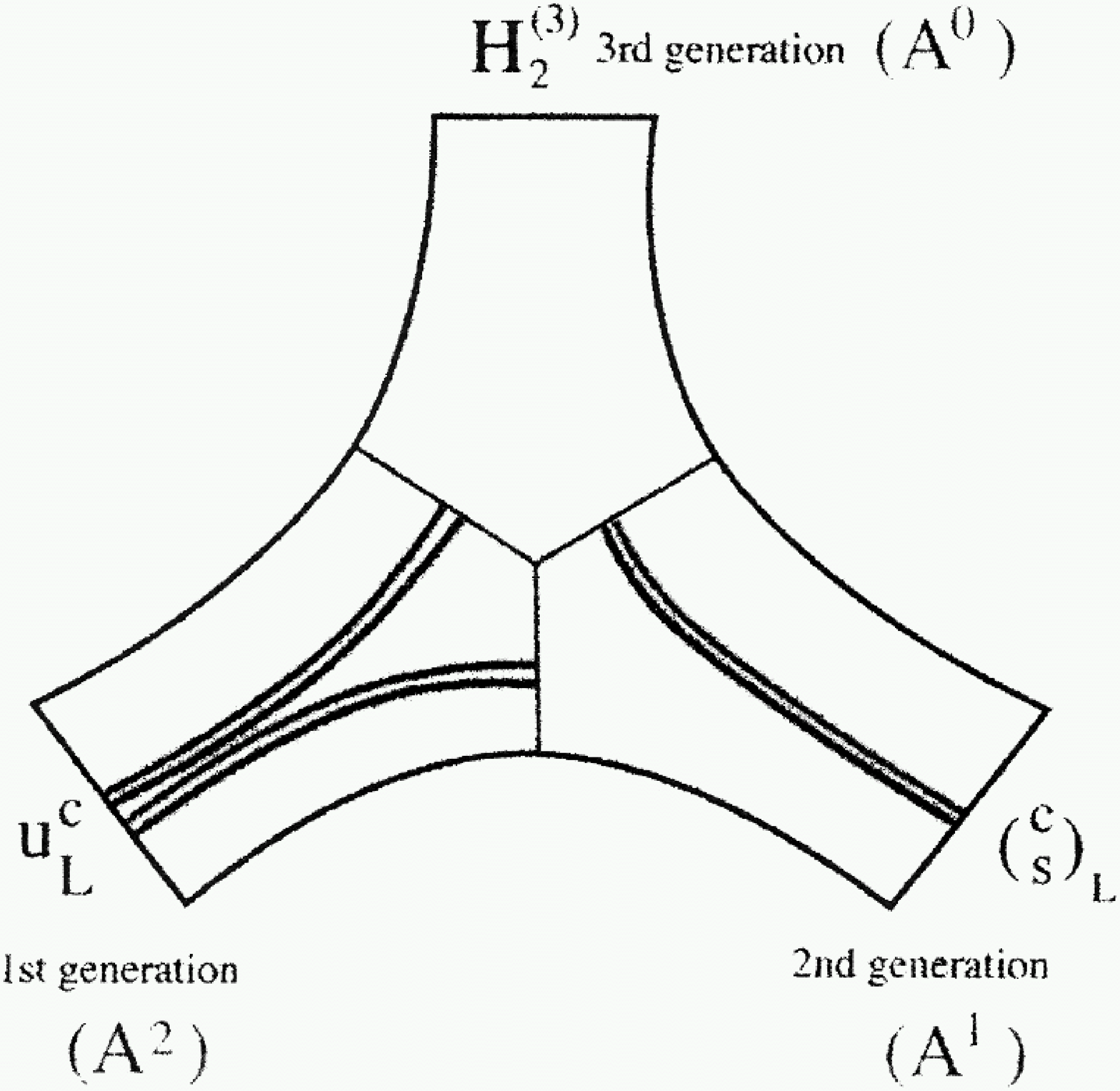, width=2in, height=3in}
\end{center}
\caption{Effective Yukawa coupling generated in the
$[Sp(6)]^3$ preon model.}
\end{figure}

These negative remarks should not obscure the considerable achievement of
these dynamical models for understanding the origin of flavor.  Families
in these models arise as a result of hidden degrees of freedom in some
underlying confining dynamics.  Furthermore, the presence of heavy excitations
in this same dynamics can result in hierarchial patterns of Yukawa couplings,
once all family symmetries are explicitly broken.  Unfortunately, it is
difficult to see how one can obtain real evidence for these kinds of
schemes, barring the discovery of some of the exotic bound states 
they predict--in the example discussed, the $T_2$ and $T_3$ states or the
vector-like partners of the quarks and leptons.

\section{Families from Short-Distance Random Dynamics}

A radically different scheme for the origin of families has been proposed
and elaborated by Holgar Nielsen and his collaborators\cite{RD}.  The 
basic idea that Nielsen has put forth is that there exist both order and
chaos at very short distances.  He imagines that at scales much smaller
than the inverse of the Planck mass there is actually a lattice structure
of scale length $a \ll 1/M_{\rm Planck}$.  However, both the dynamics on the
lattice as well as the structure of the lattice
is random.  In particular, the lattice is
amorphous with sites at random positions.  Furthermore,
characteristic of the random dynamics, the interactions on each
of the links
are governed by different groups, with the 
groups varying from link to link.

Remarkably, even starting from these very general assumptions, one can arrive
at some conclusions.  Generally, one naively 
would imagine that no group could
survive the random dynamics.  That is, that the gauge group will 
end up by breaking down
spontaneously, producing supermassive fields of mass 
$M\sim a^{-1}\gg M_{\rm Planck}$.  In fact, as Brene and Nielsen\cite{BN}
showed, there are special groups $G_{\rm surv.}$ on the links which
survive the random dynamics--i.e., the associated vector bosons are
massless.  What Brene and Nielsen\cite{BN} showed is that the  
groups which survive must have a center which is non-trivial
and connected.  By taking values in the center the links are effectively
gauge-invariant.  However, the center cannot be simply the unit matrix
because the random nature of the dynamics would then end up by averaging
out the effects of 
all links.  The connectedness of the center, finally, is necessary to
insure that the Bianchi 
identities are satisfied.  Specifically, it turns out
that $G_{\rm surv.}$ is a product of ``prime" groups with a certain discrete
group $D_{\rm prime}$, generated from the center, removed:
\begin{equation}
G_{\rm surv.} = U(1)\times SU(2)\times SU(3)\times SU(5)\times ~\ldots~
SU(\rm prime)/D_{\rm prime}
\end{equation}

From the above, it appears that Nielsen's random dynamics allows the Standard
Model group to survive, with a restriction:
\begin{equation}
G^*_{\rm SM} = SU(3)\times SU(2)\times U(1)/D_3~.
\end{equation}
Here the discrete group $D_3$ is given by powers of the center 
element $h = \left\{e^{\frac{2\pi i}{3}}I_3,-I_2,e^{\frac{2\pi i}{6}}\right\}$:
\begin{equation}
D_3 = \left\{h^n|n\subset Z_6\right\}
\end{equation}
In practice, this imposes a restriction on the matter states which are
placed on the random lattice sites
\begin{equation}
h\psi = \psi~,
\end{equation}
which fixes the hypercharge of the quarks relative to the leptons.  Eq. (9)
effectively imposes the familiar charge quantization, giving the quarks
third-integral charges.  This is a very nice result!

In this scheme the origin of family replication occurs through what Bennett,
Nielsen and Picek\cite{BNP} call ``confusion" in the random dynamic processes.
This can be understood as follows.  At some step in the random dynamics what
survives is not simply the group $G^*_{\rm SM}$ but a number $N_{\rm F}$ of
copies of $G^*_{\rm SM}$, each with one family of quarks and leptons.  
Subsequently, this product group collapses to
its diagonal subgroup $[G^*_{\rm SM}]_{\rm diag}$.  This collapse,
through ``confusion", results in $N_{\rm F}$ replicas of a Standard Model
family of quarks and leptons.  Thus, schematically, family generation
occurs in random dynamics when $N_{\rm F}$ Standard Model surviving groups
collapse:
\begin{equation}
G^*_{\rm SM}\times G^*_{\rm SM}\times~\ldots~\times G^*_{\rm SM}\to
[G^*_{\rm SM}]_{\rm diag}
\end{equation}

Bennett, Nielsen and Picek\cite{BNP} try to estimate $N_{\rm F}$--the number
of families--which arise from random dynamics confusion by making a
number of assumptions. Although
some of these assumptions are questionable, they are not unreasonable.
First, Bennett {\it et al.} suppose that the lattice scale associated with the random
dynamics is of order of the Planck scale:  $a = M_{\rm P}^{-1}$.  This
allows the calculation of the coupling constants of the Standard Model
group $[G^*_{\rm SM}]_{\rm diag}$ from
their low energy values via the renormalization group:
\begin{equation}
[g_i]_{\rm diag} \simeq g_i[M_{\rm P}]
\end{equation}
Second, by identifying the gauge fields in $[G^*_{\rm SM}]_{\rm diag}$
with the individual fields in each of the SM groups in Eq. (10), it follows
that the individual couplings of each of the groups in the ``confused"
configuration $G^*_{\rm SM}\times G^*_{\rm SM}\times ~\ldots~\times
G^*_{\rm SM}$ is given by
\begin{equation}
g_i^{\rm conf} = \sqrt{N_{\rm F}} [g_i]_{\rm diag}
\end{equation}
A knowledge of $g_i^{\rm conf}$ then provides an estimate for $N_{\rm F}$.
What Bennett, Nielsen and Picek\cite{BNP} assume is that
\begin{equation}
g_i^{\rm conf} = g_i^*~,
\end{equation}
with $g_i^*$ being the mean field theory critical coupling for each of
the groups in the Standard Model.  This assumption guarantees that in the
confusion stage there is no confinement of quark and lepton states at Planck
length scales--a reasonable boundary condition.

The result for $N_{\rm F}$ which follows from the three assumptions
(11)-(13) are rather remarkable, given the spare theoretical framework!
One finds\cite{RD}
\begin{equation}
N_{\rm F} = \left\{ \begin{array}{ll}
3.4      & {\rm U(1)} \\
3.5      & {\rm SU(2)} \\
3.1      & {\rm SU(3)}
\end{array}
\right.
\end{equation}
This result notwithstanding, however, it is not clear how one proceeds further
in developing a consistent theoretical framework from random dynamics.
For instance, it is totally unclear how through this scheme one induces the
breakdown of the $SU(2)\times U(1)$ electroweak group at scales of
$O(100~{\rm GeV})$, or how one even generates the Yukawa couplings which
can provide the quarks and leptons eventually with some mass.

\section{A Geometrical Origin for Families}

Perhaps the most interesting way to get family replications is through
the compactification of extra dimensions.  One starts with a theory in
$d>4$ dimensions but then assumes that the extra dimensions somehow
compactify, leaving a 4-dimensional theory.  The earliest example of such a
theory was the 5-dimensional Kaluza-Klein theory of gravity\cite{KK},
which when compactified to 4 dimensions gave rise, in addition to gravity,
also to electromagnetic interactions.  More modern examples are superstring
theories\cite{GSW} which are known to be consistent only in $d>4$
dimensions, but where again the extra dimensions can compactify leaving an
effective 4-dimensional theory.

It is quite easy to understand how one can generate families in these 
types of theories.  The general idea was first sketched out by
Wetterich\cite{Wetterich} and Witten\cite{Witten} in the early 1980's.
Consider chiral fermions in a d-dimensional space-time.\footnote{Chiral
fermions occur naturally in $d=2$ mod 4 dimensions.}  Such fermions, by
definition, obey a massless Dirac equation
\begin{equation}
\Gamma^\alpha D_\alpha\psi = 0
\end{equation}
Here $\alpha = 1,2,\ldots d$ and $D_\alpha$ is the Dirac operator in the
background of whatever other fields (gravity, Yang-Mills) are present in 
the theory.  Suppose now ($d-4$) dimensions compactify.  Then Eq. (15)
can be written as
\begin{equation}
(\Gamma^\mu D_\mu + \Gamma^aD_a)\psi = 0
\end{equation}
with $\mu = 0,1,2,3$ and $a = 4,~\ldots~,~d-1$.  Clearly the ($d-4$)
operator $\Gamma^aD_a$ in Eq. (16) acts as a 4-dimensional mass
$[\Gamma^aD_a\equiv M]$ {\bf unless} it vanishes when applied on $\psi$:
\begin{equation}
\Gamma_aD^a\psi = 0
\end{equation}
If Eq. (17) holds, corresponding to a chirality constraint on the 
($d-4$)-dimensional compact space $K$, then also the 4-dimensional
fermions will be chiral.  If (17) does not hold, then the resulting
4-dimensional fermions have a mass.

Since the quarks and leptons are chiral, if they are produced from $d>4$
chiral fermions via compactification of the extra dimension, a constraint
equation like (17) on the compact space $K$ must hold.  Now, in general,
such constraint equations have a number of solutions,\footnote{Think of
solving a differential equation in a periodic box.} which depend on the
intrinsic properties of the compact space $K$.  So, in these kinds of
theories, families and family number are intrinsically related to the
topological properties of compact spaces associated with the original $d>4$
theory.

Perhaps the best known example of family replication using these ideas is
the one considered by Candelas, Horowitz, Strominger, and Witten\cite{CHSW}
involving the Calabi-Yau compactification of the $d=10$ heterotic superstring.
This string theory\cite{heterotic} has an associated $E_8\times E_8$
gauge symmetry and is supersymmetric.  The chiral fermions in the $d=10$
theory are gauginos of one of the $E_8$ groups (the other $E_8$ acts as
a hidden sector), sitting in the 248 dimensional adjoint 
representation.\footnote{Majorana fermions exist in $d=2$ mod 8 dimensions.}
Candelas {\it et al.}\cite{CHSW} assumes that the 10-dimensional
space of the theory compactifies down to $d=4$ Minkowski space times a
6-dimensional Calabi-Yau space, whose principal property for our purposes
is that it possesses an $SU(3)$ holonomy.  This means that the chiral zero
modes in $K$--those that obey the constraint equations (17)--have 
non-trivial $SU(3)$ properties,
even though this $SU(3)$ is broken in the compactification. 
By decomposing
$E_8$ into its $E_6\times SU(3)$ subgroup one identifies the chiral zero
modes in the gauginos which are candidates for the surviving chiral matter
in 4 dimensions.  Since, under this decomposition,
\begin{equation}
248 = (78,1)\oplus (27,3)\oplus (2\bar 7,\bar 3) + (1,8)
\end{equation}
one sees that, after Calabi-Yau compactification, the 4-dimensional chiral
matter involve fermions in either the 27 or $2\bar 7$ reprentations of
$E_6$.  So, in general, one expects to have $n_{\rm F}$ 27 plus 
$\delta(27 + 2\bar 7)$ states in the spectrum.  The numbers $n_{\rm F}$ 
and $\delta$ are related to topological indices characteristic of the 
Calabi-Yau space $K$ which compactified.  In particular, Candelas
{\it et al.}\cite{CHSW} showed that $n_{\rm F}$--the number of families--is
connected to the Euler number of $K$:
\begin{equation}
n_{\rm F} = \frac{1}{2}~ n_{\rm Euler}
\end{equation}

Note that, in this example, the families one obtains have the right stuff.
The 27-dimensional representation of $E_6$ when decomposed in terms of its 
$SO(10)$ subgroup contains the 16-dimensional representation, appropriate
for a family of quarks and leptons, plus a 10 and a singlet.  The 10
itself, since the theory is supersymmetric, contains the two needed Higgs
doublets, which in this case also come in family repetitions.  In
principle, the $\delta(27 + 2\bar 7)$ states (as well as the 10 and 1) are
vectorlike, and one can imagine these states getting masses of the order of
the compactification scale--presumably of $O(M_{\rm P})$.  So in this
example, the light states are just $n_{\rm F}$ replications of the chiral
quarks and leptons!

Connecting family replication to the geometry of a compact space is a
beautiful idea.  Furthermore, there is another advantage.  Through
compactification, Yukawa couplings are naturally produced, arising from the
fermion-gauge field interactions in $d>4$ dimension along the gauge field
components in the $(d-4)$ compact dimensions.  Unfortunately, however, one
cannot in general compute these couplings explicitly.  Nevertheless, often one
can infer some useful symmetry restrictions among the Yukawa 
couplings in these schemes\cite{GKR}.

In my view, obtaining families from compactification is the most 
appealing solution to the origin of the mysterious repetitions we see in
nature.  It is not, however, easy to arrive at the correct theory.  Basically,
even believing that superstrings are the right theory,
we still do not understand how to choose among the many possible 
compactifications available for these theories, since we have no idea
of what is the underlying physics principle that drives the compactification.
At the same time, we are also ignorant of how these schemes can give rise to
terms which break supersymmetry and eventually $SU(2)\times U(1)$.
Until such problems are solved, these ideas will just remain ideas which are
appealing but untested.

\section{Navigating through the Mass Maze}

Even if one were to eventually understand the origin for families and their
matter content, the mystery of flavor will not be solved until one is
able also to decipher the physics which leads to the peculiar mass spectrum
of quarks and leptons.  Lacking a complete theory, most physicists have taken
a very pragmatic approach to the mass generation problem.  Basically, what
has been assumed is that this problem is essentially decoupled from that
of families and matter.  Therefore, it makes sense to pursue a quite
phenomenological strategy to get some insights into the problem of mass.

Following the lead set by some early work of Weinberg\cite{Weinberg} and
Fritzsch\cite{Fritzsch}, the strategy has been to assume that the mass
matrices for the fermions have certain ``textures", imposed on them by
some underlying symmetries.  These textures, in turn, allow one to derive
some interesting ``predictions" which can then be compared with experiment.
Typically, one obtains in this way certain interrelations among the
quark mixing angles and the quark masses--relations which go
beyond the standard model.

Perhaps the most famous ``prediction" of this type of approach is the 
following formula for the Cabibbo angle, expressed as a function of quark mass ratios:
\begin{equation}
\sin\theta_c = \sqrt{\frac{m_d}{m_s}} - \sqrt{\frac{m_u}{m_c}}
\end{equation}
Equation (20) follows directly, in the case of two generations, if the quark
mass matrices have the Fritzsch pattern\cite{Fritzsch}
\begin{equation}
M_u = \left(
\begin{array}{cc}
O & A \\
A & B
\end{array}
\right) ~; ~~~~
M_D = \left(
\begin{array}{cc}
O & C \\
C & D
\end{array}
\right)
\end{equation}
which display a texture zero in the 11 matrix element.  Given that Eq. (20)
is rather successful phenomenologically, the natural question to ask in
this context is the underlying reason for the appearance of the texture
zero in Eq. (21).

The appearance of texture zero or other interrelations between the elements
in the quark and lepton mass matrices are generally assumed to arise at some high scale $M$ where new physics connected with mass generation comes into
play.  In models where the breakdown of $SU(2)\times U(1)$ is dynamical
like Technicolor\cite{TC}, the scale $M$ is generally assumed to be not too
far from the TeV scale.  However, in general, one has to be careful with
FCNC induced through the process of mass generation\cite{DE} and one
must appeal to dynamical properties of the underlying theory\cite {WTC}
to avoid contradiction with experiment.  The resulting theories are quite
complicated\cite{Holdom} and, as a result, many physicists think it more
likely that the scale $M$ connected with mass generation is likely to be of
order of the Planck or GUT scale.  In what follows, I shall concentrate only
on this latter possibility and discuss two different, but complementary, mechanisms
which can provide mass matrices with interesting textures: family symmetries
and GUTs.

I will illustrate the first of these possibilities by discussing briefly a
model introduced by Iba\~nez and Ross\cite{IR}, which makes use of a
$U(1)_{\rm F}$ family symmetry.\footnote{In the literature, there are many
models which use a $U(1)$ symmetry as a family symmetry\cite{Nir}, starting
from the original paper on flavor textures by Froggatt and Nielsen\cite{FN}.}
In the Iba\~nez-Ross model, the quarks and antiquarks of each generation
have the opposite $U(1)_{\rm F}$ charge, while the two Higgs bosons
$H_1$ and $H_2$ of the model carry twice the $U(1)_{\rm F}$ charge of the
third generation:
\begin{equation}
\left(
\begin{array}[c]{c}
u \\ d
\end{array}
\right)^i_{\rm L} \to
e^{i\alpha_i}
\left(
\begin{array}{c}
u \\ d
\end{array}
\right)^i_{\rm L}
;
\begin{array}[c]{c}
\left(u^c_L\right)^i \to e^{-i\alpha_i}\left(u^c_L\right)^i \\
\left(d^c_L\right)^i \to e^{-i\alpha_i}\left(d^c_L\right)^i
\end{array}
;
\begin{array}[c]{c}
H_1 \to e^{2i\alpha_3}H_1 \\
H_2 \to e^{2i\alpha_3}H_2
\end{array}
\end{equation}
As a result of this symmetry, clearly the quark mass matrices only
have a non-zero 33 element:
\begin{equation}
M_{\rm 0}^{u,d} = m_{t,b}
\left(
\begin{array}{ccc}
0 & 0 & 0 \\
0 & 0 & 0 \\
0 & 0 & 1
\end{array}
\right)
\end{equation}
This provides a reasonable starting point for model building.

To proceed, Iba\~nez and Ross\cite{IR} need to introduce both a way to
break the $U(1)_{\rm F}$ symmetry and some interactions which physically
will serve to generate the integenerational mass splittings.  They
accomplish the second point by imagining that at some high scale $M$
some $SU(2)\times U(1)$ singlet fields $\theta$ and $\bar\theta$, with
$U(1)_{\rm F}$ charges of +1 and -1, respectively, acquire some effective
interactions with the quark and Higgs fields.  How these effective
interactions come about need not be specified, but the $U(1)_{\rm F}$
symmetry will fix their form.  Iba\~nez and Ross essentially make use of
the Froggatt-Nielsen\cite{FN} mechanism we illustrated earlier with the
$[Sp(6)]^3$ preon model.  For example, there will be a $U(1)_{\rm F}$
preserving effective interaction among the $u$ and $t$ quarks and the two
scalar fields $H_2$ and $\theta$ of the form
\begin{equation}
{\cal{L}}_{ut}^{\rm eff} \sim
\left(\frac{\theta}{M}\right)^{\alpha_1-\alpha_3} ~~
\bar u_{\rm L} H_2 t_{\rm L}^c
\end{equation}

Iba\~nez and Ross\cite{IR}, in addition, assume that the $U(1)_{\rm F}$
family symmetry is itself spontaneously broken at high scales by VEVs of
the $\theta$ and $\bar\theta$ fields.  Once $\theta$ acquires a VEV a
term like (24) becomes an effective Yukawa interaction, which can give
rise to small corrections to the mass matrix (23) if 
$\epsilon = \frac{\langle\theta\rangle}{M} \ll 1$.  Of course, within this
approach, one is not able to predict precisely these corrections.  Nevertheless,
if one assumes that the proportionality constants in the Froggatt-Nielsen
terms are of $O(1)$, the magnitude of the different matrix elements will be
governed by powers of $\epsilon$, reflecting the original $U(1)_{\rm F}$
symmetry.  In the case of the Iba\~nez and Ross model, for example, the
modified up-quark mass matrix under these assumptions takes the 
form\footnote{One can take $\alpha_1+\alpha_2+\alpha_3 = 0$, without loss of
generality}
\begin{equation}
M_1^u \sim m_t
\left(
\begin{array}{ccc}
\epsilon^{|-4\alpha_1-2\alpha_2|} &
\epsilon^{|-3\alpha_1|} &
\epsilon^{|-\alpha_2-2\alpha_1|} \\
\epsilon^{|-3\alpha_1|} &
\epsilon^{2|\alpha_2-\alpha_1|} &
\epsilon^{|\alpha_2-\alpha_1|} \\
\epsilon^{|-\alpha_2-2\alpha_1|} &
\epsilon^{|\alpha_2-\alpha_1|} &
1
\end{array}
\right)
\end{equation}
Not only is this mass matrix hierarchial if $\epsilon \ll 1$,
but there are interesting interrelations among the matrix elements;
for example
\begin{equation}
(M_1^u)_{11} \simeq \frac{\left[(M_1^u)^2_{13}\right]^2}
{(M_1^u)_{33}}~; ~~~~
(M_1^u)_{22} \simeq \frac{\left[(M_1^u)_{23}\right]^2}
{(M_1^u)_{33}}
\end{equation}

In general, the detailed comparison of a mass matrix like that in Eq. (25),
which holds at a large scale $M$, with experiment is complicated by the
evolution of the Yukawa couplings with energy.  This evolution, for
example, can change zeros in a mass matrix at a given scale into small, but
non-vanishing, contributions at a different scale.  This is easy to
understand since, for example, a non-vanishing Yukawa coupling to quarks
of the second generation can induce at one loop an effective Yukawa
coupling to quarks of the first generation, provided there is also a
non-vanishing Yukawa coupling between the first two generations.

The effect of the renormalization group evolution of the couplings is to
further obscure possible mass matrix patterns.  For instance, the mass
matrix $M^u$ of Eq. (25) is connected to the Yukawa couplings $\Gamma_u$
via the familiar equation $M^u = \frac{\Gamma_u}{\sqrt{2}}\langle H_2\rangle$.
However, $\Gamma_u$ at one scale is different from $\Gamma_u$ at another
scale, with the evolution between scales governed by the renormalization
group.  At one loop, this evolution is given by the equation:
\begin{eqnarray}
16\pi^2 \frac{d \Gamma_u}{d \ln M/\mu}   = & 
\left\{{\rm Tr}
~(3\Gamma_u\Gamma_u^{\dagger} + 3a\Gamma_d\Gamma_d^{\dagger} +
a\Gamma_\ell\Gamma_\ell^{\dagger}) \right. \nonumber\\
& \left. -G_u^2 + \frac{3}{2}
(b\Gamma_u\Gamma_u^{\dagger} + c\Gamma_d\Gamma_d^{\dagger})\right\} \Gamma_u
\end{eqnarray}
In the above, the coefficients $a,b,c$ and $G_u^2$ depend on the assumptions
one makes on the matter content between the scale $M$ and the scale $\mu$.
As a result, patterns set by physics at a high scale $M \sim O(M_{\rm Planck})$
at lower energies $\mu$ are smudged.  In particular, the measured mass
matrix at $\mu \sim M_{\rm Z}$ is influenced by the assumptions one makes
on the matter content in the region between $\mu$ and $M$--a region for
which one has no information!  Nevertheless, once one fixes this matter
content through some model assumptions, one can either validate or exclude
specific mass matrix models by comparing their, renormalization group
evolved, predictions with precision electroweak data.

Rather than illustrating this for the model of Iba\~nez and Ross sketched
above, I prefer instead to examine the predictions of a specific class of
SUSY GUT models, based on $SO(10)$ studied by Anderson, Dimopoulos, Hall,
Raby and Starkman\cite{ADHRS}.  These models illustrate a second way in which
theoretical input at a high scale may help fix the patterns of the quark and
lepton masses and mixing we observe.   These SUSY $SO(10)$ models
have texture zeros and matrix element interrelations set by $SO(10)$, with
hierarchies among these elements produced by insertion of higher dimensional
operators much along the lines of what transpired in the Iba\~nez and Ross
model.
The best model of \cite{ADHRS} with three texture zeros has the following
Yukawa couplings:
\begin{eqnarray}
\Gamma_u &=&
\left(
\begin{array}{ccc}
0 & C & 0 \\
C & 0 & -\frac{B}{3} \\
0 & -\frac{4B}{3} & A
\end{array}
\right) ~; ~~~~
\Gamma_d =
\left(
\begin{array}{ccc}
0 & -27C & 0 \\
-27C & Ee^{i\phi} & \frac{B}{9} \\
0 & -\frac{4B}{9} & A
\end{array}
\right) ~; \nonumber \\
\Gamma_\ell &=&
\left(
\begin{array}{ccc}
0 & -27C & 0 \\
-27C & 3Ee^{i\phi} & B \\
0 & 2B & A
\end{array}
\right)
\end{eqnarray}
The parameters $A,B,C$ and $E$ have the following hierarchy
\begin{equation}
A \gg B \gg E > C
\end{equation}
and the model Clebsch-Gordan coefficients have a Georgi-Jarlskog\cite{GJ}
pattern for $\Gamma_{22}:~ [u;d;\ell] = [0;1;3]$, which produces the
nice result
\begin{equation}
\frac{m_s}{m_d} \simeq \frac{m_\mu}{9m_e}
\end{equation}
Although these and other results provide an acceptable fit to the 
low-energy data, a recent analysis by Blacek, Carena, Raby and 
Wagner\cite{BCRW} has shown that these models do not fit the data
as well as models with just one texture zero, like the Lucas-Raby
model\cite{LR}.

The Lucas-Raby model adds two non-zero matrix elements to the Yukawa
coupling of Eq. (28), with strength $D\sim O(C)$.  Specifically one has:
\begin{eqnarray}
(\Gamma_u)_{13} = -\frac{4D}{3} e^{i\delta}~~&; 
(\Gamma_u)_{31} = -\frac{D}{3} e^{i\delta}~~&;
(\Gamma_d)_{13} = \frac{2D}{3} e^{i\delta}~~; \nonumber \\
(\Gamma_d)_{31} = -9De^{i\delta}~~&;
(\Gamma_\ell)_{13} = -54De^{i\delta}~~&; 
(\Gamma_\ell)_{31} = -De^{i\delta}
\end{eqnarray}
Table 1
displays a comparison of the fits provided by these two models of all
the extant low-energy data.  As can be seen, perhaps not surprisingly,
the data clearly favors the Lucas-Raby model with all predictions within
one standard deviation from the data.  In contrast, the best of the 
Anderson {\it et al.}\cite{ADHRS} models has four observables about
$3\sigma$ away from the data:
\hskip3cm\begin{table}
\begin{center}
\begin{tabular}{|c|c|c|l|c|l|c|}
\cline{1-3} \cline{5-5} \cline{7-7} 
Observable & Central Value & $\sigma$ & 
\hskip1cm & Result of\cite{ADHRS} & \hskip1cm &
Result of\cite{LR} \\
\cline{1-3} \cline{5-5} \cline{7-7} 
$M_\tau$ & 175.0 & 6.0 & & 173.9 & & 175.7 \\
$m_b(M_t)$ & 4.26 & 0.11 & & 4.360 & & 4.287 \\
$M_b-M_c$ & 3.4 & 0.2 & & 3.146 & & 3.440 \\
$m_s$ & 180 & 50 & & 162.6 & & 189.0 \\
$m_d/m_s$ & 0.05 & 0.015 & & 0.0461 & & 0.0502 \\
$Q^{-2}$ & 0.00203 & 0.00020 & & 0.00173 & & 0.00204 \\
$M_\tau$ & 1.777 & 0.0089 & & 1.777 & & 1.776 \\
$M_\mu$ & 105.66 & 0.53 & & 105.6 & & 105.7 \\
$M_e$ & 0.5110 & 0.0026 & & 0.5113 & & 0.5110 \\
$V_{us}$ & 0.2205 & 0.0026 & & 0.2215 & & 0.2205 \\
$V_{cb}$ & 0.0392 & 0.003 & & 0.0450 & & 0.0400 \\
$V_{ub}/V_{cb}$ & 0.08 & 0.02 & & 0.0463 & & 0.0772 \\
$B_K$ & 0.8 & 0.1 & & 0.9450 & & 0.8140 \\
\cline{1-3} \cline{5-5} \cline{7-7} 
\end{tabular}
\end{center}
\caption[]{Comparison of the Models of Ref.\cite{ADHRS} and Ref.\cite{LR}
with Experiment: Adapted from\cite{BCRW}}
\end{table}
$Q^{-2} = \frac{m_d^2-m_u^2}{m_s^2}~; ~~~ V_{cb}~; ~~
V_{ub}/V_{cd} ~~\hbox{and} ~~ \hat B_K$ 
--the parameter characterizing the strength of the $K^o-\bar K^o$ matrix
element which enters in the CP violating $\epsilon$ parameter.

I should remark that, besides fitting the extant data, these models 
are {\bf predictive}.  Once all the parameters (A-E and the phases) in
Eqs. (28) and (31) are fixed from the global fit, one can extract further
information from these ansatze, both on the CKM matrix as well as on the
spectrum of Higgs and supersymmetric states.  For example,
the Lucas-Raby model predicts the following values for the parameters
associated with the CKM unitary triangle, so important for studies of CP
violation in the B system\cite{LR}.
\begin{equation}
\sin 2\alpha = 0.96~; ~~ \sin 2\beta = 0.52~; ~~~\sin\gamma = 0.93~;
\rho = -0.125~; ~~~\eta = 0.32
\end{equation}
The angles $\alpha,\beta$ and $\gamma$ will be soon measured and Eq. (32)
will be tested experimentally.  At the same time, the Lucas-Raby model also
predicts that the lightest Higgs bosons are a pair of CP odd and CP even
states nearly degenerate with each other with mass around 74 GeV--a prediction
which should be tested already by LEP 200.

\section{Lessons from the Top Mass.}
Because $m_t \gg m_i$, many of the important features of the Yukawa matrices
are crucially dependent on how the top coupling behaves.  Theoretically,
rather than considering the physical mass for the top measured by the
CDF and DO Collaborations\cite{PDG} $M_t = (175 \pm 6)$ GeV, it is more useful to consider
the running mass\footnote{We use the same convention as Table 1 in which
lower case masses are running masses and capital case masses are physical
(pole) masses.}
\begin{equation}
m_t(M_t) \sim
\frac{M_t}{1+\frac{4}{3\pi} \alpha_s(m_t)} =
(167 \pm 6)~{\rm GeV}
\end{equation}
The running mass is directly related to the diagonal Yukawa coupling of the
top $\lambda_t(m_t)$
\begin{equation}
m_t(m_t) = \lambda_t(m_1)\langle H_2\rangle
\end{equation}
This coupling, keeping only the dominant 3rd generation
couplings in Eq. (27) obeys
the RG equation\cite{OP}
\begin{equation}
\frac{d\lambda_t(\mu)}{d\ln \mu} =\frac{1}{(4\pi)^2}
\left\{a_t\lambda_t^2(\mu) + a_b\lambda_b^2(\mu) + a_\tau\lambda_\tau^2(\mu) 
 -4\pi c_i\alpha_i(\mu)\right\}\lambda_t(\mu)
\end{equation}
The coefficients $a_i,c_i$, as we discussed earlier, depend on the
matter content of the theory.  For instance, for the SM one has
\begin{equation}
a_t = \frac{9}{2}~;~~
a_b = \frac{3}{2}~;~~
a_\tau = 1~;~~
c_i = \left(\frac{17}{20}~;~~
\frac{9}{4}~;~~8\right)~,
\end{equation}
while for the minimal supersymmetric extension of the Standard Model (MSSM)
one has
\begin{equation}
a_t = 6~;~~a_b = 1~;~~ a_\tau = 0~;~~
c_i = \left(\frac{13}{5}~;~~3~;~~\frac{16}{3}\right)
\end{equation}

Knowing the value for $\lambda_t(m_t)$ one can compute the value for
$\lambda_t(\mu)$ at any scale $\mu$ by using the RG equation (35).  Because
$a_t > 0$, for $\mu$ large enough eventually $\lambda_t(\mu)\to\infty$.
The location of this, so called, Landau pole\cite{Landau} is 
theory-dependent.  As we shall see, $\mu_{\rm Landau}$ is an uninteresting
scale in the SM, but for the MSSM $\mu_{\rm Landau} \sim M_{\rm Planck}$--a
result which may be quite significant.  At any rate, it is worthwhile to
explore these differences in a bit more detail\cite{SW}.

For the Standard Model, because one has only one Higgs boson, one has
$\langle H_2\rangle \equiv \langle H\rangle \simeq \frac{1}
{2^{3/4}G_F^{1/2}} \simeq 174~{\rm GeV}$ and $\lambda_t(m_t) = 0.96 \pm 0.04$.
Furthermore, in this case, because top is so heavy $\lambda_t^2 \gg
\lambda_b^2~,~\lambda_\tau^2$ and these other couplings can be safely
dropped from Eq. (35).  The solution of the RG equation
\begin{equation}
\frac{d\lambda_t(\mu)}{d\ln \mu}=
\frac{1}{(4\pi)^2}
\left\{a_t\lambda_t^2(\mu)-4\pi c_i\alpha_i(\mu)\right\}
\lambda_t(\mu)
\end{equation}
is easily found to be
\begin{equation}
\lambda_t^2(\mu) =
\frac{\eta(\mu)\lambda_t^2(m_t)}
{\left[1 - \frac{a_t}{8\pi^2}\lambda^2_t(m_t)I(\mu)\right]}~.
\end{equation}
Here $\eta(\mu)$ and $I(\mu)$ are functions determined by the running of
the gauge coupling constants
\begin{equation}
\frac{d \alpha_i(\mu)}{d\ln \mu}=
\frac{b_i}{2\pi} \alpha_i^2(\mu)
\end{equation}
with $b_i \equiv \left(\frac{41}{10}~,~~ -\frac{19}{6}~, ~~ -7\right)$ and one finds
\begin{equation}
\eta(\mu) = \Pi_i\left[\frac{\alpha_i(m_t)}{\alpha_i(\mu)}\right]^{c_i/b_i}~;
I(\mu) = \int^{\ln(\mu)}_{\ln(m_t)} \eta(\mu^\prime) d\ln\mu^\prime.
\end{equation}
From Eq (39) one sees that the Landau pole for the SM model occurs at a value of $\mu$ where
\begin{equation}
I(\mu_{\rm Landau})\biggr|_{\rm SM} =
\frac{8\pi^2}{a_t\lambda_t^2(m_t)} \simeq 18.9
\end{equation}
Using Eq. (36) one finds $\mu_{\rm Landau} \simeq 10^{32}$ GeV a scale well
beyond the Planck mass, which has clearly no physical significance.  At the
Planck mass, $I(M_{\rm Planck}) \sim 12.2$, so one is still far away from the
Landau pole.  Indeed $\lambda_t(M_{\rm Planck}) \sim 0.7 < 
\lambda_t(m_t)$.  So the top Yukawa coupling in the SM remains perturbative
$(\lambda_t^2/4\pi \ll 1)$ up to the Planck scale.

The situation is quite different if there is supersymmetry.  Because there
are now two Higgs doublets, $\lambda_t(m_t)$ is not fixed by $m_t(m_t)$ but
depends also on the ratio of the two Higgs VEV, $\tan \beta$.  One has
$\langle H_2\rangle = \frac{1}{2^{3/4}G_F^{1/2}}\sin\beta$, so that
\begin{equation}
\lambda_t(m_t) = \frac{0.96\pm 0.04}{\sin\beta}
\end{equation}
There are two interesting regions for $\tan\beta$.  The first of these has
$\tan\beta\sim O(1)$, in  which case one can again neglect $\lambda_b$ with
respect to $\lambda_t$ in Eq. (35).  In the second region
$\tan\beta \gg 1$, but $\lambda_t(m_t)\simeq \lambda_b(m_t)$.  That is, there
is a unification of the top and bottom Yukawa couplings.  In either of these
two regions Eq. (35) reduces to the same approximate form (38).  However, now
one has either
\begin{equation}
a_t = 6~; ~~~ \lambda_t(m_t) = \frac{0.96\pm 0.04}{\sin\beta}
\end{equation}
if $\tan\beta \sim O(1)$.  If $\tan\beta \gg 1$ instead, one has
\begin{equation}
a_t = 7~; ~~~ \lambda_t(m_t) = 0.96 \pm 0.04 
\end{equation}

The form of the solution for $\lambda_t^2(\mu)$ in these two cases is again
given by Eq. (39).  However, here the running of the gauge couplings which
enter in $\eta(\mu)$ and $I(\mu)$ is governed by a different set of
coefficients $b_i$ and $c_i$, appropriate to having supersymmetric matter
above the weak scale.  In particular, $b_i = \left(\frac{33}{5}~,~~1~,~~-3
\right)$.  Given Eqs. (44) and (45), it is clear that the Landau pole will
occur at the same place in both cases, provided that in the first case
$\sin\beta = \sqrt{\frac{6}{7}}~(\tan\beta \simeq 2.45)$.  In what follows,
therefore, we concentrate only on the second case, corresponding to Yukawa
unification of couplings.

The Landau pole in this case occurs at
\begin{equation}
I(\mu_{\rm Landau})\biggr|_{\rm MSSM} =
\frac{8\pi^2}{a\lambda^2_t(m_t)} = 12.15\pm 1.00
\end{equation}
Using the MSSM coupling evolution, it is easy to check that such values for
I obtain when $\mu_{\rm Landau} \sim O(M_{\rm Planck})$.  Indeed, the error
on $\lambda_t$ is large enough not to permit a more accurate determination.
In fact, it is much more sensible to turn the argument around.  If there is
supersymmetry and $\lambda_t$ becomes strong around $M_{\rm Planck}$, then
at low energy $\lambda_t$ will be driven to an infrared fixed point at
$\lambda_t(m_t) \simeq 1$\footnote{At one-loop level, this fixed point occurs
at a value of $\lambda_t$ where the RHS of Eq. (38) vanishes:
$\lambda^*_t = \lambda_t(m_t) =
[64\pi \alpha_3(m_t)/21]^{1/2} \simeq 1.03$.}
This is illustrated by Fig. 2, calculated using the two-loop MSSM RGE
equation for $\tan\beta = 2$, where the ``focusing" effect at low scales
of couplings which are strong around $\mu = M_{\rm Planck}$ is clearly
demonstrated.

\begin{figure}
\begin{center}
\epsfig{file=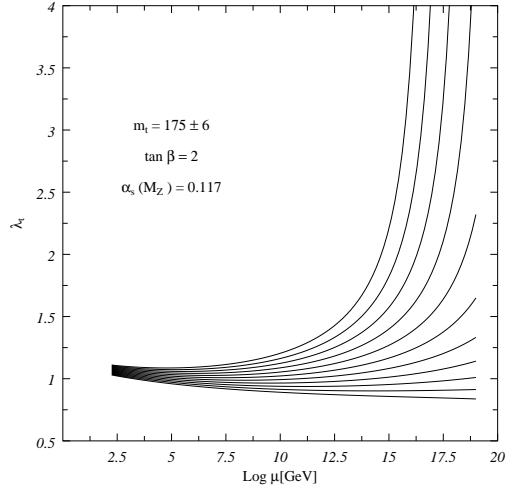,height=3in}
\end{center}

\caption{Focusing of the Yukawa couplings as $\lambda_t\to\lambda_t^*$.}
\end{figure}

The results displayed in Fig. 2 are perfectly consistent with having the
ratio $m_b/m_\tau = 1$, as SO(10) unification suggests, at scales of
$O(M_{\rm Planck})$.  An analysis similar to the one we did for Eq. (38) relates
this ratio at a scale of $m_t$ to that at the unification scale
$M_X \sim M_{\rm Planck}$:
\begin{equation}
\frac{m_b(m_t)}{m_\tau(m_t)} =
\frac{m_b(M_X)}{m_\tau(M_X)}
\left[\frac{\eta(M_X)}{\hat\eta(M_X)}\right]^{1/2}
\left[\frac{\lambda_t(M_X)}{\lambda_t(m_t)}\right]^{-1/6}~.
\end{equation}
Here $\hat\eta(\mu)$ is a quantity similar to $\eta(\mu)$, detailing the
running of the coupling constants in the quark to lepton mass ratio:
\begin{equation}
\hat\eta(\mu) = \Pi_i
\left(\frac{\alpha_i(m_t)}{\alpha_i(\mu)}\right)^{\hat c_i/b_i}
\end{equation}
with the coefficients $\hat c_i = \left(-\frac{4}{3}~,~~0~,~~\frac{16}{3}
\right)$ for the MSSM.  The ratio $[\eta(M_X)/\hat\eta(M_X)]^{1/2} \simeq
1.68$ is above the experimental ratio \linebreak  
$m_b(m_t)/m_\tau(m_t) =
1.58\pm 0.08$, suggesting that $\lambda_t(M_X) \gg \lambda_t(m_t)$.  That is,
the top coupling is stronger at the GUT scale than at low energy, much
as indicated in Fig. 2.

The upshot of this discussion is that the assumption that 
supersymmetric matter exists above the
weak scale gives a consistent picture, with a large top Yukawa coupling at
the Planck scale being driven by an infrared fixed point to a value
$\lambda_t(m_t) \sim 1$.  This behavior obtains in two regimes of $\tan\beta$.
Either $\tan\beta \sim O(1)$ and $\lambda_t$ is the dominant coupling.
Or $\lambda_b \sim \lambda_t$ and $\tan\beta$ is large.  The second
possibility is natural in the SO(10) models discussed earlier where all quarks
and leptons of one family are in the 16-dimensional representation.
Furthermore, at least intuitively, having a large Yukawa coupling at the 
Planck scale fits in well with the ideas that families are generated either
dynamically or through geometry in supersymmetric theories.

\section{Concluding Remarks}

In my opinion, one probably will not be able to unravels the mystery of
flavor {\bf without} some new experimental information.  In particular, I
believe that ascertaining whether or not low energy supersymmetry exists
will have a profound impact on this question.  The discovery of low energy
supersymmetry would, of course, provide a tremendous boost for superstring
theories.  At the same time, it would also signal the death knell of the
random dynamics ideas of Nielsen.  These ideas, if one is to believe in them,
require that there should be a real desert up to $M_{\rm Planck}$, with no
physics beyond the Standard Model between the weak and the Planck scale.

If supersymmetry is found, perhaps it is sensible to imagine that some of
the ideas discussed in the previous section are true.  That is, that there
is indeed a large Yukawa coupling of top at energies of $O(M_{\rm Planck})$,
which results in the mass of the top being determined essentially by the 
infrared fixed point of the Yukawa evolution equations.  Furthermore, 
it is easy to imagine then that the
quark and lepton mass spectrum is a result of a combination of a broken
family symmetry--which sets up the hierarchy among the masses--and of a
GUT--which interrelates the quark and lepton mass tapestries.

Even in this very favorite circumstance, however, it will be difficult to
get real evidence for the origin of flavor.  Is it due to dynamics or to
some primordial compactification?  Perhaps the tell-tale sign will emerge
from the discovery of some exotic states, besides the quarks and leptons and
their superpartners.  In fact, the most characteristic signals of models for
flavor is the inevitable presence of exotic states.  Recall the exotic $T_2$
and $T_3$ states in the $Sp(6)$ model, or the extra $10 \oplus 1$ states in
the 27 produced through a Calabi-Yau compactification.  In this respect,
I should note that certain exotic states seem to be quite generic.  In
particular, the presence of extra $(3,1)_{-1/3}$ states is very natural.

On a more pedestrian level, our undestanding of flavor and mass will be
aided by a continuous experimental (and theoretical) refinement of the
values for the quark and lepton masses and mixing parameters.  Precise values
for these parameters are crucial if one wants to sort out alternative 
tapestries, signalling different origins for flavor.  Eventually, it is going
to be important to know that $V_{cb} = 0.038$ rather than 0.040!

{\bf Acknowledgements}

I am grateful to Ikaros Bigi and Luigi Moroni for their hospitality at the
Varenna School.  A condensed version of these lectures was presented in
Chicago, Illinois, at the Symposium ``20 Years of Beauty Physics", while a
more popular version was given as the first Abdus Salam Memorial Lecture in
Islamabad, Pakistan.  I thank Dan Kaplan and Ahmed Ali, respectively, for their
kind invitations.  
This work is supported in part by
the Department of Energy under Grant \# DE-FOO3-91ER40662, Task C.

\end{document}